\documentstyle[psfig]{mn}
\def\Sref#1{Section~\ref{#1}}

\def\altaffilmark#1{$^{#1}$}
\def\altaffiltext#1#2{\\$^{#1}${#2}}
\def\plotone#1{\centerline{\psfig{figure=#1,width=\hsize}}}
\def\acknowledgements{\section*{Acknowledgements}}
\makeatletter
\def\ion#1#2{#1$\;${\small\rm\@Roman{#2}}\relax}
\def\@cite#1#2{#1\if@tempswa , #2\fi}
\let\cite\@internalcite
\makeatother
\def\Sref#1{\S\ref{#1}}
\def\G{G\,}
\def\GD{GD\,}
\def\zzpsc{\G29-38}
\def\RV{\ifmmode R_{\rm V}\else\hbox{$R_{\rm V}$}\fi}
\def\RC{\ifmmode R_{\rm C}\else\hbox{$R_{\rm C}$}\fi}
\def\DV{\ifmmode\Delta\Phi_{\rm V}\else\hbox{$\Delta\Phi_{\rm V}$}\fi}
\def\DC{\ifmmode\Delta\Phi_{\rm C}\else\hbox{$\Delta\Phi_{\rm C}$}\fi}
\def\kms{\ifmmode{\rm km\,s^{-1}}\else\hbox{$\rm km\,s^{-1}$}\fi}
\def\lobs{\ifmmode\lambda_{\rm obs}\else$\lambda_{\rm obs}$\fi}
\def\msun{\ifmmode M_\odot\else$M_\odot$\fi}
\def\l{\ifmmode\ell\else$\ell$\fi}
\def\m{\ifmmode m\else$m$\fi}
\def\n{\ifmmode n\else$n$\fi}
\def\d{{\rm d}}

\begin{document}

\title[Mode identification in the pulsating white dwarf \zzpsc]{Mode
identification from time-resolved spectroscopy of the pulsating white
dwarf \zzpsc}

\author[J. C. Clemens, M. H. van Kerkwijk and Y. Wu]{%
        J. C. Clemens\altaffilmark{1,2},
        M. H. van Kerkwijk\altaffilmark{1,3,4} and
        Y. Wu\altaffilmark{5,6}
\altaffiltext{1}{Palomar Observatory, California Institute of
                 Technology 105-24, Pasadena, CA 91125, USA}
\altaffiltext{2}{Department of Physics and Astronomy, University of
                 North Carolina, Chapel Hill, NC 27599-3255, USA;
                 clemens@physics.unc.edu}
\altaffiltext{3}{Institute of Astronomy, University of Cambridge,
		 Madingley Rd, Cambridge, CB3 0HA, UK} 
\altaffiltext{4}{Astronomical Institute, Utrecht University, P.O. Box
                 80000, 3508 TA Utrecht, The Netherlands;
                 M.H.vanKerkwijk@astro.uu.nl}
\altaffiltext{5}{Theoretical Astrophysics, California Institute of 
		 Technology 130-33, Pasadena, CA 91125, USA}
\altaffiltext{6}{Astronomy Unit, School of Math.\ Sci.,
		 Queen Mary and Westfield College, Mile End Road,
		 London E1 4NS, UK; Y.Wu@qmw.ac.uk}
}

\maketitle

\begin{abstract}
  We have used time-resolved spectroscopy to measure the colour
dependence of pulsation amplitudes in the DAV white dwarf \zzpsc.
Model atmospheres predict that mode amplitudes should change with
wavelength in a manner that depends on the spherical harmonic degree
\l\ of the mode.  This dependence arises from the convolution of mode
geometry with wavelength-dependent limb darkening.  Our analysis of
the six largest normal modes detected in Keck observations of \zzpsc\
reveals one mode with a colour dependence different from the other
five, permitting us to identify the \l\ value of all six modes and to
test the model predictions.  The Keck observations also show pulsation
amplitudes that are unexpectedly asymmetric within absorption lines.
We show that these asymmetries arise from surface motions associated
with the non-radial pulsations (which are discussed in detail in a
companion paper).  By incorporating surface velocity fields into line
profile calculations, we are able to produce models that more closely
resemble the observations.
\end{abstract}


\section{Introduction}\label{sec:intro}

Despite observations over a broad range of wavelengths and via
numerous techniques, the star \zzpsc\ remains an enigma.  It is the
third brightest ZZ~Ceti known ($V=13.05$), and has pulsation
amplitudes among the largest measured for these variables (up to 6\%
modulations at optical wavelengths). It is the most extensively
observed large amplitude ZZ~Ceti star, having been the subject of two
global observing campaigns (Winget et al.\ \cite{wing&a:90}; Kleinman
et al.\ \cite{klei&a:94}) and of countless single-site time series
measurements (Kleinman et al.\ \cite{klei&a:98}).  It has also been
the target of several infrared and radial velocity studies following
the detection of an infrared excess by Zuckerman \& Becklin
(\cite{zuckb:87}).  Nevertheless, we still have neither an
unambiguous asteroseismological solution for the star, nor an
explanation for the source of the mysterious dust apparently
responsible for its excess emission in the infrared.

  Recently, Kleinman (\cite{klei:95}, also Kleinman et al.\
\cite{klei&a:98}) analyzed all of the optical time-series photometry
of this object, and showed that in spite of the changing character of
the pulsation spectrum each season, there is a stable set of recurring
modes.  This is an important breakthrough for \zzpsc, and perhaps for all
the large amplitude ZZ~Ceti stars, because measuring mode periods is a
prerequisite for measuring mass and internal structure using stellar
seismology.  The only remaining obstacle is to identify the spherical
harmonic degree \l\ and radial order \n\ of the modes detected, so
that their periods can be compared to those of like eigenmodes in
structural models of white dwarf stars.

   The pattern identified by Kleinman (\cite{klei:95}) is sufficiently
rich that he was able to attempt mode identification using the same
techniques applied successfully to DOV (Winget et al.\
\cite{wing&a:91}) and DBV pulsators (Winget et al.\ \cite{wing&a:94}).
He searched for the (roughly) equal period spacings and frequency
splittings that signify rotationally split non-radial {\it g}-modes.
His attempt was a measured success; he found that the pattern of modes
was sensible if interpreted as a sequence of mostly $\l=1$ modes.
Unfortunately, his analysis could not assure that any individual mode
was $\l=1$, nor did it allow an unambiguous comparison to structural
models (Bradley \& Kleinman \cite{bradk:96}).

  With this in mind, and having available to us the bright portion of
a night allocated to very faint sources at the Keck II telescope, we
decided to test a method for identifying the degree of pulsation modes
using \zzpsc\ as our subject.  The method was inspired by the work of
Robinson et al.\ (\cite{robi&a:95}), who used Hubble Space Telescope
high speed photometry in the ultraviolet to measure \l\ for modes in
the star \G117-B15A.  Their technique exploited the sensitivity of
ZZ~Ceti mode amplitudes to the wavelength of the observations.  At all
wavelengths, the observed amplitudes are diminished by cancellation
between surface regions with opposite pulsation phase.  In the
ultraviolet, stronger limb darkening changes the character of this
cancellation and observed mode amplitudes differ from their optical
values in a manner that depends on \l.  The models calculated by
Robinson et al.\ (\cite{robi&a:95}) to explore differences between
ultraviolet and optical also show amplitude changes within optical
absorption lines.  The character of these changes likewise depends
upon \l, providing the potential to determine \l\ from optical
spectroscopy.

  In an attempt to measure the line profile variations of \zzpsc\ and
use them for \l\ identification, we acquired over four hours of
time-resolved spectroscopy using the Keck II Low Resolution Imaging
Spectrometer (Oke et al.\ \cite{oke&a:95}).  We have described these
observations in a companion paper (Van Kerkwijk et al.\
\cite{vank&a:99}, Paper~I), and presented an analysis of the
periodicities present in the total flux and line-of-sight velocity
curves.  Prior to our Keck observations, the velocity variations
associated with ZZ~Ceti pulsations had never been detected.  Their
presence significantly complicates the models required to fully
understand our data, but also increases the amount of valuable
information we can hope to extract.

  In this paper we present our analysis of the line profile variations
of \zzpsc.  We begin in \Sref{sec:spec}, with an analysis of the
average spectrum, which has a signal to noise ratio higher than usual
for white dwarf spectra.  In \Sref{sec:obs} we present the amplitudes
and phases as a function of wavelength for the largest modes and
compare them to each other and to theoretical models like those
calculated by Robinson et al.\ (\cite{robi&a:95}).  On this basis
alone it will be clear that we can unambiguously identify \l\ for
these modes.  It will also be clear that good quantitative fits to the
data will require improvements to model atmospheres.  Nonetheless, we
will continue in \Sref{sec:vel} by incorporating the velocity field
associated with the pulsations into the models.  The improved models
help to constrain other pulsation properties, such as the velocity
amplitude of motions at the stellar surface.  We present our
conclusions in \Sref{sec:conc}.

\section{The Mean Spectrum}\label{sec:spec}

By averaging together all of our time-series spectra, we have
constructed a mean spectrum with extremely high signal to noise ratio.
This spectrum shows Balmer lines (H$\beta$ through H$\iota$), the
\ion{Ca}{2}~$\lambda3933$ resonance line, and a hint of
\ion{Mg}{2}~$\lambda4481$ (Paper~I).  Metal lines were first
discovered in \zzpsc\ by Koester, Provencal, \& Shipman
(\cite{koesps:97}), who also found iron lines in the ultraviolet
spectrum.  The presence of metals in a DA spectrum is unusual, and
Koester et al.\ (\cite{koesps:97}) attribute them to the accretion of
dust, supporting the notion that the infrared excess in the spectrum
of \zzpsc\ is caused by reprocessing of light by circumstellar dust
grains.

We have fitted the Balmer lines in our mean spectrum using a grid of
model spectra kindly provided by D. Koester (for a recent description,
see Finley, Koester, \& Basri \cite{finlkb:97}).  The models consist
of tabulated intensities, $I_\lambda$, at 9 limb angles, $\mu =
cos(\theta)$, for a grid of atmospheres with effective temperatures
spanning the ZZ~Ceti instability strip and with gravities from $\log g
= 7.50$ to 8.75. The atmospheres were all calculated using the ML2,
$\alpha = 0.6$ prescription for convection, which yields consistent
fits for ZZ Ceti stars over the broadest range of wavelengths
(Bergeron et al.\ \cite{berg&a:95}).  To compare the models to our
spectrum we integrated $I_\lambda$ over the visible hemisphere and
compared the resulting Balmer lines to those we observed by
normalising both model and data to fixed continuum points.  This
method is similar to that used by Bergeron, Saffer \& Liebert
(\cite{bergsl:92}), but less sophisticated than the procedure Bergeron
et al.\ (\cite{berg&a:95}) used for their analysis of ZZ~Ceti spectra.

\begin{figure}
\plotone{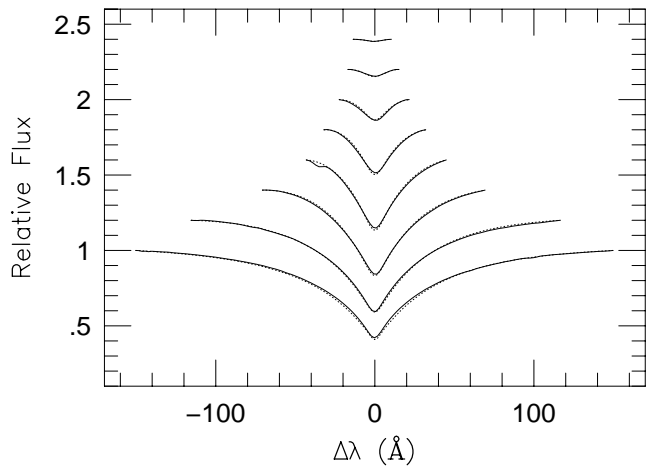}
\caption[]{The best overall fit to the line profiles H$\beta$ through
H11 in the average spectrum of \zzpsc.  The model ({\it dotted line})
has $\log g = 8.05$ and $T_{\rm eff} = 11,850$~K. The dip on the blue
side of H$\epsilon$ is the \ion{Ca}{2}~$\lambda3933$ line.
\label{fig:specfit}}
\end{figure}

  Figure~\ref{fig:specfit} shows the observed Balmer lines along with
the best fitting model, which has $T_{\rm eff} = 11,850$~K and $\log g
= 8.05$.  These values are close to those published for \zzpsc\ by
Bergeron et al.\ (\cite{berg&a:95}) and Koester et al.\
(\cite{koesps:97}).  The former found $T_{\rm eff} = 11,820$~K and
$\log g = 8.14$, the latter $T_{\rm eff} = 11,600$~K and $\log g =
8.05$.  Using the evolutionary models of Wood (\cite{wood:94}) with
thick surface H layers ($\sim 10^{-4} M_*$) our values translate into
a mass of 0.64 \msun.

As impressive as the fit in Figure~\ref{fig:specfit} is, the
discrepancies between model and data are still dominated by systematic
effects, rather than stochastic noise.  This makes the values of
$\chi^2$ we calculate useless for evaluating the error in our
temperature and gravity determination; the error is dominated by real
differences between the models and the measurements. These differences
might arise from a variety of sources. 

One possibility is that the normalisation of our data to the continuum
points was affected by errors in our calibration of the instrumental
response.  Another possibility is that errors arise from the presence
of metal lines not included in the models.  The shape of $\rm
H8=H\zeta$ and H$\epsilon$ are probably affected by depression of the
intervening continuum by the \ion{Ca}{2}~$\lambda3933$ line and
H$\epsilon$ is contaminated by \ion{Ca}{2}~$\lambda3968$.  However,
the metal lines cannot entirely account for the problem, because the
fits to H$\beta$, H$\gamma$, and H$\delta$, which should be unaffected
by metals, are also less than perfect.

  The discrepancy in the fits might also be due to a problem discussed
by Koester, Allard, \& Vauclair (\cite{koesav:94}).  They found that
even the best prescription for convection yields a temperature
structure in model atmospheres that is only approximately correct.
Consequently, the synthetic spectra the models produce will not match
observed spectra at every wavelength simultaneously nor will they
match all the line profiles, which are highly sensitive to the run of
temperature with depth.  To explore this possibility, and to see how
our final fit is affected by individual lines, we have fitted separate
models to each of the Balmer lines.

\begin{figure}
\plotone{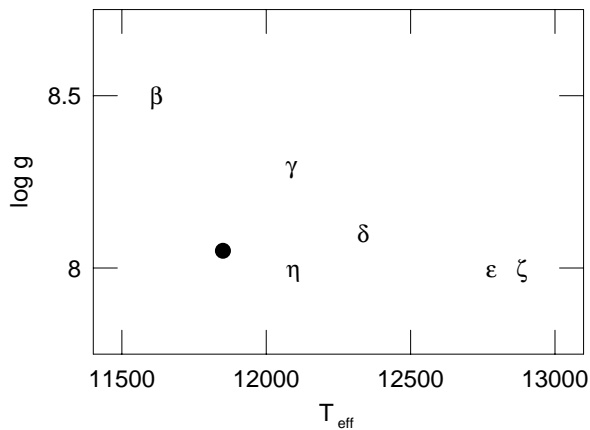}
\caption[]{Location of best fits to the individual Balmer lines in the
$\log g$--$T_{\rm eff}$ plane.  Fits to individual lines are denoted
by the line designations; the best fit using all the lines (see
Figure~\ref{fig:specfit}) is shown as a filled circle.
\label{fig:linefit}}
\end{figure}

In Figure~\ref{fig:linefit}, we plot the values of $\log g$ and
$T_{\rm eff}$ for the best-fitting models of the individual lines
H$\beta$ through H9.  They span a range of temperature from 11,620 to
12,885~K.  With the exception of $\rm H9=H\eta$, the fits show a trend
to higher $T_{\rm eff}$ and lower $\log g$ as the excitation level
increases.  The location of the fit using all lines compared to the
individual lines shows that the gravity fit is dominated by the lines
of higher excitation, which are known to be more gravity sensitive.
Conversely, the temperature is fixed by the lower excitation lines,
which change more rapidly with $T_{\rm eff}$.

Figure~\ref{fig:linefit} is not meant to suggest that the individual
fits give answers inconsistent with the global fit, rather they should
be regarded as helping to establish the uncertainty of the temperature
and gravity determination.  It is clear that high signal-to-noise
ratio alone is not enough to improve the determination of temperature
and gravity. Better calibration of the instrument or improvements to
the models (or both) will be necessary.  The systematic trend in the
individual Balmer line fits suggests the problem lies with the models,
as do the results of the following section. There we compare
fractional amplitudes calculated from data and models, and find
further discrepancies.  In the fractional amplitudes, calibration
errors should cancel to first order.

   If the models turn out to be responsible for the discrepancies we
measure, our high signal to noise average spectrum will provide the
opportunity to improve atmospheric models, perhaps even to infer the
correct temperature profile.  Furthermore, individual spectra in our
time series, which themselves have a signal to noise ratio of 100,
span a temperature range of about 500~K and will show how the
atmospheric structure should change with model temperature.  This
information about the derivative of the thermal profile should provide
extremely valuable constraints on model atmospheres.  Our data are
available upon request to anyone interested in exploring these
problems.

   The final factor that might affect our model fit is the presence of
relatively large amplitude pulsations.  We have shown in Paper~I that
the velocity fields associated with the pulsations of \zzpsc\ are
detectable in our data.  These motions alter the average spectrum,
mainly by Doppler broadening the absorption lines very slightly.  The
models we will describe in \Sref{sec:vel} allow us to incorporate this
effect into the model spectrum.  Fitting the data with these velocity
broadened spectra yields a slight improvement in the quality of the
fit to the average spectrum, but an insubstantial change in the values
of $\log g$ and $T_{\rm eff}$ we infer from the models.

\section{Line Profile Variations}\label{sec:obs}

  Robinson et al.\ (\cite{robi&a:95}) have described and implemented a
method for distinguishing the value of \l\ in ZZ~Ceti stars by
comparing pulsation amplitudes at different wavelengths.  Their method
relies upon the increased importance of limb darkening in the
ultraviolet.  Non-radial pulsation modes of every \l\ suffer from
geometric dilution in their amplitudes when averaged over the visible
hemisphere.  This dilution increases with \l\ as cancellation between
regions with opposing phase becomes important.  However, at short
wavelengths increased limb darkening diminishes the effect of the
cancellations.  For $\l \leq 3$, the net result is that mode
amplitudes increase in the ultraviolet relative to their optical
values in a way that depends on~\l.  Robinson et al.\
(\cite{robi&a:95}) used optical and ultraviolet high speed photometry
to measure this effect for the 215~s pulsation mode in \G117-B15A.
They were able to conclude that this mode is $\l=1$.  Fontaine et al.\
(\cite{font&a:97}) reanalyzed the same data using independent models
and arrived at the same conclusion, although they differed with
Robinson et al.\ in the model temperature that best fits the data.

\begin{figure}
\plotone{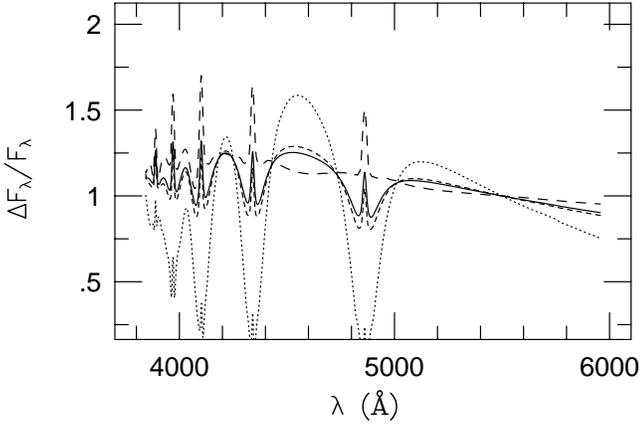}
\caption[]{Models for the wavelength dependent flux variations for
modes of $\l = 1$ through 4, after Robinson et al.\
(\cite{robi&a:95}).  We have used the values of $\log g$ and $T_{\rm
eff}$ from our best fit to the average spectrum and convolved the
output with a Gaussian of 5 \AA\ width to match our Keck observations.
All curves are normalised at 5500~\AA.  The solid line is $\l = 1$,
the short dashed line is $\l=2$, the dotted line is $\l = 3$, and the
long dashed line is $\l = 4$
\label{fig:robmodels}}
\end{figure}

The models Robinson et al.\ (\cite{robi&a:95}) calculated also showed
\l-dependent differences in the behaviour of pulsation amplitudes
within the absorption lines.  These too arise from the effects of limb
darkening on modes of different \l.  Figure~\ref{fig:robmodels} shows
these changes for low values of $\l$, which are expected to dominate
the modes observed in white dwarfs.  We calculated these curves using
a modified version of code originally provided to us by E. L. Robinson
(see Robinson et al.\ \cite{robi&a:95}).  Instead of integrating over
a limb darkening law, we integrate over intensities tabulated for
different values of $\mu$, as described in \Sref{sec:spec}.
Consequently, Equations~3a and 3b in Robinson et al., which represent
the equilibrium flux and the flux changes due to the pulsations,
become:
\begin{equation}
F_\lambda = 2\pi R_0^2\int_0^1 I_\lambda(g,T_0,\mu) \mu\; \d\mu,
\label{eq:fl0}
\end{equation}
and
\begin{eqnarray}
\Delta F_\lambda&=&2\pi R_0^2 (R_0 \frac{\delta T}{\delta R})\epsilon
k_{\l\m} \cos ( \sigma t) \times \nonumber \\
&&\int_0^1\left.\frac{\partial I_\lambda (g,T,\mu)}{\partial T}\right|_{T_0}
 P_\l(\mu)
\mu\; \d\mu.
\label{eq:dfl}
\end{eqnarray}
Here, $R_0$ is the equilibrium stellar radius, $T_0$ the equilibrium
temperature, $\epsilon$ is the amplitude of the radius changes induced
by the pulsations, and $\frac{\delta T}{\delta R}$ is the Lagrangian
derivative of temperature with respect to radius. Together
$P_{\l}(\mu)$ and $k_{\l m}$ represent the surface distribution of the
temperature changes after integrating in the $\phi$ direction.
$P_{\l}(\mu)$ is a Legendre polynomial depending only on $\mu$, and
$k_{\l m}$ depends on the angle between the pulsation axis and the
observer's line of sight.  Our notation is slightly different from
Robinson et al.\ (\cite{robi&a:95}) in that we have separated the time
dependence, $\cos (\sigma t)$, from $k_{\l m}$.  Finally, while
$I_\lambda(g,T,\mu)$ comes directly from the tabulated models,
$\frac{\partial I_\lambda(g,T,\mu)}{\partial T}$ must be calculated by
taking differences between models of different $T_{\rm eff}$.

Robinson et al. (\cite{robi&a:95}) have emphasized the useful
properties of the ratio $\Delta F_\lambda/F_\lambda$ for \l\
identification.  It is not sensitive to the flux calibration of the
data and varies with wavelength in a way that does not depend on mode
inclination and \m\ (see also the Appendix).  In
Figure~\ref{fig:robmodels} we show $\Delta F_\lambda/F_\lambda$ for
modes of $\l=1$ through 4.  The curves have been normalised to one at
5500 \AA.  We have not included $\l = 0$, which resembles $\l = 1$
with slightly smaller modulations, because it is clear from the long
periods of the modes that they cannot be radial pulsations.

\begin{figure}
\plotone{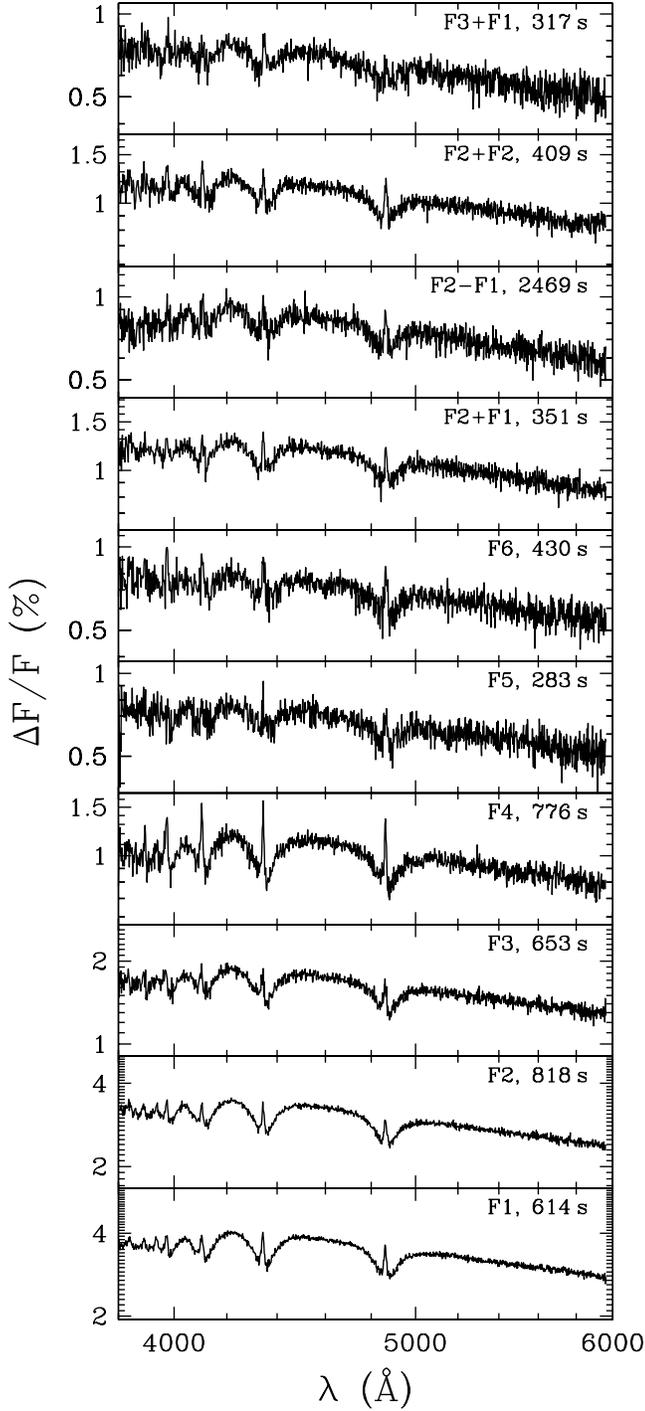}
\caption[]{Wavelength dependent amplitudes for the 6 largest modes and
4 largest combination frequencies in \zzpsc.  To ease comparison, the
same logarithmic scale was used for all panels.  Note the small peak
seen in the amplitudes of F1 at 3933~\AA, corresponding to the
\ion{Ca}{2} line.
\label{fig:amps}}
\end{figure}

In order to compare our data to the models in
Figure~\ref{fig:robmodels}, we have fitted the amplitudes and phases
of the 11 largest pulsation frequencies in each 2 \AA\ wavelength bin
using the function $A\cos(2\pi{}ft-\phi)$.  During these fits, we held
the mode frequency, $f$, constant at the values tabulated in Paper~I,
and fit the 11 amplitudes and phases simultaneously.  Of these 11
modes, five are combination frequencies, i.e., frequencies which have
values that are sums or differences of larger modes.  We have
discussed the nature of combination frequencies in Paper~I and will
return to them at the end of this section.  Figure~\ref{fig:amps}
shows the fractional amplitudes at each wavelength for the six
physical modes and for the four largest combination frequencies.  The
fifth, F3$-$F1 at 10,322\,s, we judged too noisy to include.  The
qualitative similarity between the data and the models is striking.
Even in the line cores the models predict the behaviour of mode
amplitudes quite well, despite the fact that within $\sim\!1\,$\AA\ of
the core non-LTE effects in the atmosphere are important, which are
not included in the model.  We emphasize that these are predictions in
the literal sense; Koester calculated the atmosphere models before we
acquired the data.

\begin{figure}
\plotone{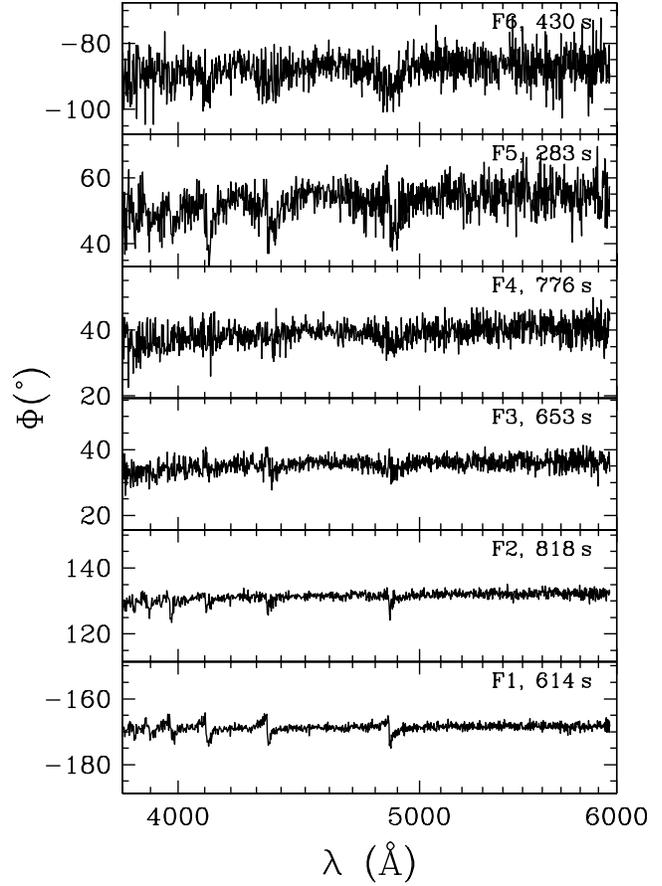}
\caption[]{Wavelength dependent phases for the 6 largest normal modes in
\zzpsc.
\label{fig:phase}}
\end{figure}

In the models we have discussed so far, the pulsations have the same
phase at every wavelength.  However, the phases of the physical modes,
shown in Figure~\ref{fig:phase}, show distinct changes in the vicinity
of spectral features.  For mode F1, the phase changes within
absorption lines bear the signature of a velocity induced variation;
they change with the derivative of the spectrum.  We will return to
these phase changes in \Sref{sec:vel}, when we have models capable of
reproducing them.  In addition to the phase changes within the lines,
Figure~\ref{fig:phase} also shows a small slope in the continuum
phases, indicating that pulse maximum in blue light arrives earlier
than in red by a few seconds.  This slope is not reproduced by our
models, and remains a mystery.

Careful inspection of the amplitudes of the real modes in
Figure~\ref{fig:amps} reveals that those for mode F4, at 776~s, show a
different shape than for the other modes; they increase more sharply
in the line cores and curve more steeply in the continuum.  F4 was
already noticeably different from the other real modes in Paper~I,
where we found it had a larger velocity to light amplitude ratio than
any of the other modes, and it produced a stronger harmonic.

\begin{figure}
\plotone{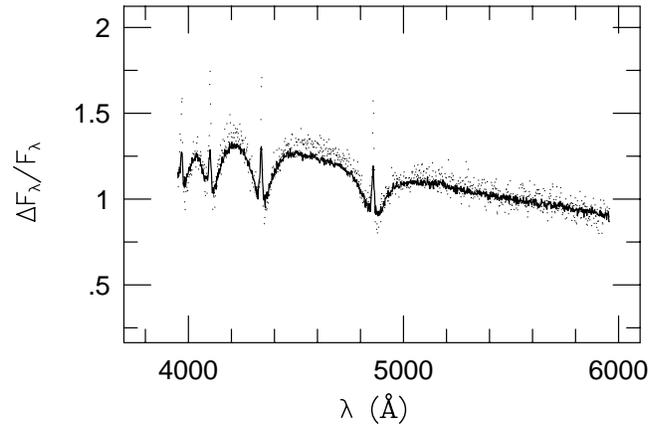}
\caption[]{Amplitude comparison for F1 ({\it solid line}) and F4 ({\it
discrete points}).  The amplitudes have been normalised to 1 at
5500~\AA.
\label{fig:14comp}}
\end{figure}

To help see the differences in F4 we have normalised the amplitudes of
F1 and F4 at 5500~\AA\ and plotted the modes together in
Figure~\ref{fig:14comp}.  The most likely explanation for the
differences we see is that F4 has a value of \l\ different from F1.
Qualitative application of the theoretical models demands that \l\
must be higher for F4, to yield the larger changes in amplitude we
observe.  Furthermore, the large contrast between modes of $\l = 2$
and higher in the models, versus the more modest differences seen in
Figure~\ref{fig:14comp}, point toward the conclusion that F4 is $\l =
2$ and F1 is $\l=1$.

\begin{figure}
\plotone{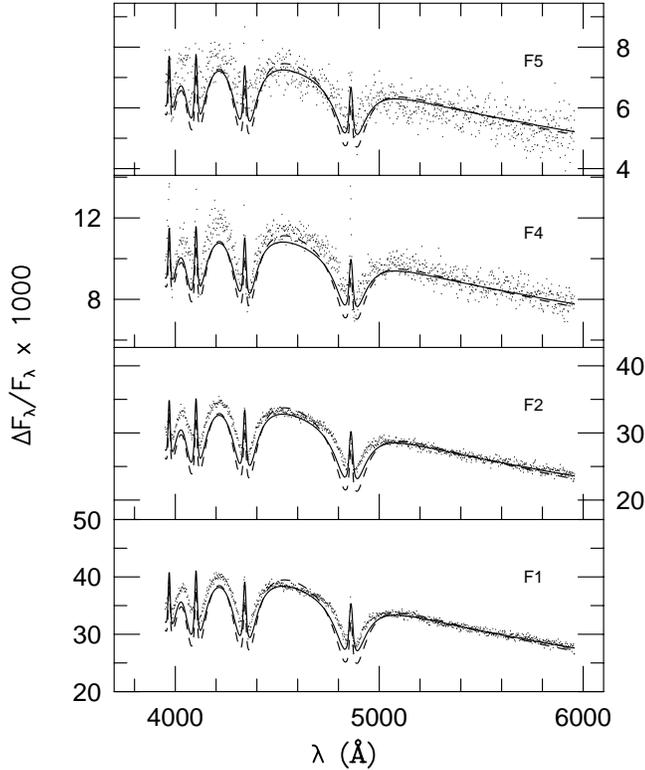}
\caption[]{Comparison to models for modes 1, 2, 4, and~5.  The solid
line is $\l = 1$, the dashed line is $\l = 2$.  The models were
calculated using $\log g = 8.05$ and equilibrium $T_{\rm eff} =
11850$~K, as derived from the fits to the average spectrum.
\label{fig:ampfit}}
\end{figure}

  To test this possibility further, we can make a direct comparison
between the data and models.  In Figure~\ref{fig:ampfit} we have
plotted the theoretical and observed amplitudes for modes 1, 2, 4,
and~5.  To establish the normalisation, we multiplied the theoretical
curves by the amplitudes of each mode at 5500~\AA.  We have used a
model with gravity and equilibrium temperature inferred from our fit
to the average spectrum in \Sref{sec:spec}.  As with the average
spectrum, Figure~\ref{fig:ampfit} shows discrepancies between the data
and the models.  The slope of the amplitude changes is steeper in the
data than in the models, worsening the fit at short wavelengths.
Correcting this would require hotter models at shorter wavelengths,
the same trend required to fit to individual Balmer lines in the
average spectrum.  We could not find a model at any single temperature
that offered a substantially better fit than the model we have used.

  In spite of the difficulties with the fits, it is clear that $\l=1$
is a better match to modes F1, F2, and F5 while $\l=2$ is a better fit
to mode~F4.  Interestingly, this is most apparent in the continuum
variations between 4500 and 4700~\AA; the models are too poor a fit
within the lines to provide a measure of \l\ there.  Nonetheless, our
original expectation that the line profiles would offer the most
sensitive \l\ discriminant are borne out by the large differences
between the amplitudes at line center in F4 and those in the other
modes.  We have included F5 as the representative of the noisier low
amplitude modes.  Even with the higher noise it is clear that $\l=1$
is a better fit to this mode, as it is to every mode except mode~F4.

Even though the quantitative agreement is poor at many wavelengths,
the presence of two apparently different values of \l\ makes our
identification secure.  The values of \l\ chosen are independent of
the model temperature.  Lowering the effective temperature of the
model chosen diminishes the changes in mode amplitudes, so that at
some $T$, the $\l = 2$ model would fit mode F1 and those like it, but
at that temperature, no value of \l\ fits F4.  Likewise, attempts to
fit F4 with $\l=1$ by increasing model temperature leave no \l\ of
lower value to fit the other modes. Consequently, on the basis of
qualitative behaviour alone, we can conclude that the 776~s mode is $\l
= 2$, and the remaining 5 modes are $\l = 1$.

\begin{figure}
\plotone{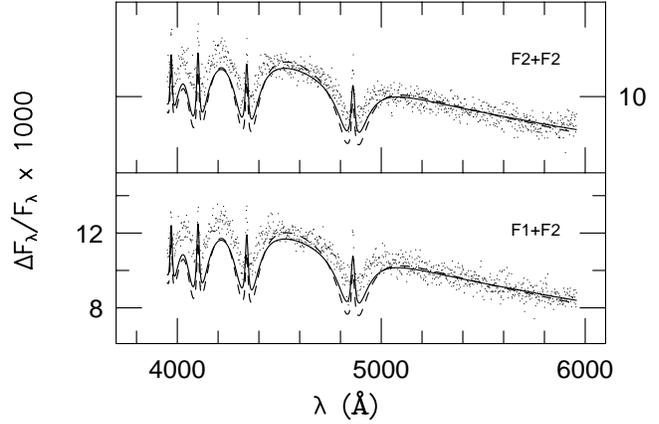}
\caption[]{Comparison to models for frequencies F1+F2
and F2+F2.  The solid line is $\l = 1$, the dashed line is $\l = 2$.
\label{fig:combo}}
\end{figure}

Finally, we consider the combination modes shown in
Figure~\ref{fig:amps}.  The combination frequencies we see in \zzpsc\
and other ZZ~Ceti stars probably do not arise from eigenmode
pulsations, but from a non-linear transformation of the modes in the
outer layers of the white dwarf (Brickhill \cite{bric:92},
cf. Brassard, Fontaine, \& Wesemael \cite{brasfw:95}).  This mixes the
modes present, generating signals at sums and differences of the mode
frequencies.  The $\l$ character of these modes depends on the $\l$ of
the modes which produce them.  For example, we expect the combination
of two $(\l,\m)=(1,0)$ modes to have $\l = 0$ and $\l = 2$ components,
while those of two $(1,-1)$ modes or two $(1,1)$ modes should produce
only $l=2$.  These expectations arise from the mathematical properties
of spherical harmonics only. It is impossible to make quantitative
predictions about combination modes without a detailed theory
explaining how they are produced.  We have shown two combination
frequencies along with models in Figure~\ref{fig:combo}.  Like the
other combinations, they most resemble the modes we have identified as
$\l=1$.  We have already discussed the implications of this in
Paper~I.

\section{Models Including Velocities}\label{sec:vel}

 Apart from the problems with quantitative fits to the model
atmospheres, there is a significant qualitative difference between the
observed and model amplitudes in Figure~\ref{fig:ampfit}; the observed
amplitudes are asymmetric within the absorption lines, while the model
amplitudes are not.  We showed in Paper~I that the Balmer lines not
only show changes in flux during a pulsation cycle, but also changes
in line-of-sight velocity.  In that paper, we treated these as
separable components of the spectral variations, but in reality they
are components of a more complex line profile variation.  In
anticipation of improvements in the model fits, in this section we
will calculate these variations by incorporating velocities into the
flux integrals we calculate at each wavelength.  This requires that we
follow a more general treatment than that used by Robinson et al.\
(\cite{robi&a:95}) to produce Equation~\ref{eq:fl0}.  Our development
will rely upon Robinson, Kepler, \& Nather (\cite{robikn:82}) and use
the same notation where possible.

   The effect of the velocity field on our calculations of the
integrated flux is that the values of $I_\lambda$ we look up from the
tabulated spectra, and the values of $\frac{\partial
I_\lambda}{\partial T}$ we calculate from them, must be chosen using
wavelengths adjusted for the velocity of each surface element.  Thus
$I_\lambda$ in Equation~\ref{eq:fl0}, and $\frac{\partial
I_\lambda}{\partial T}$ in Equation~\ref{eq:dfl} are now functions of
velocity.  Since the velocity has a time dependence, both quantities
are now also implicit functions of time.  We may write the total flux
at each observed wavelength as a function of time as:

\begin{equation}
F(\lobs,t) = F_{T_0}(\lobs) + \Delta F_T(\lobs,t),
\label{eq:flobs}
\end{equation}
where the $T_0$ is a reminder that the first term comes from the model
at equilibrium temperature, and the $T$ indicates that the flux
changes in the second term are due to changes in temperature, which
dominate all other sources (Robinson et al.\ \cite{robikn:82}).  It
will help in understanding Equation~\ref{eq:flobs} to think about two
different limits.  The first is a (hypothetical) pulsation mode which
has surface motions but no temperature changes.  For these modes, the
second term would be 0.  The first term would be a function of time
for any $\lobs$ near spectral features that could be Doppler shifted
in and out of $\lobs$ by the changing velocities.  The second limit to
think of is a mode with temperature but no velocity variations, then
the two terms in Equation~\ref{eq:flobs} reduce to the expressions given by
Equations~\ref{eq:fl0} and~\ref{eq:dfl}.  An important feature of
Equation~\ref{eq:flobs} that is not present in Equations~\ref{eq:fl0}
and~\ref{eq:dfl} is the possibility for the phase of flux maximum to
differ from the phase of temperature maximum.  This means the time of
flux maximum can differ with wavelength, a possibility that was never
allowed by Equation~\ref{eq:fl0}.

Our expressions for the terms in Equation~\ref{eq:flobs} must also be
more general than before. We can write $F_{T_0}$ generally as
\begin{equation}
F_{T_0}(\lobs) = R_0^2\int_0^{2\pi}\int_0^1 I(g,T_0,\mu,\lambda)\;\mu\;
\d\mu\; \d\phi,
\label{eq:fl0obs}
\end{equation}
which is analogous to the expression used by Kepler (\cite{kepl:84})
in his discussion of line profile variations due to {\it r}-mode
pulsations.  As we have noted, the time dependent velocities enter
into the wavelength $\lambda$, so $F_{T_0}(\lobs)$ is a function of
time.  Likewise, $\Delta F_T$ is given by
\begin{eqnarray}
\Delta F_T(\lobs,t)&=&(R_0 \frac{\delta T}{\delta r})\epsilon R_0^2 {\rm e}^{-i \psi}
\times \nonumber \\
&&\int_0^{2\pi}\!\!\int_0^1
\left.\frac{\partial I(g,T,\mu,\lambda)}{\partial T}\right|_{T_0}
\xi_r\; \mu\; \d\mu\; \d\phi, \nonumber \\
\;
\label{eq:dflobs}
\end{eqnarray}
where $\xi_r$ is the assumed functional form for the perturbations in
stellar radius,
\begin{equation}
\xi_r = Y_{\ell m}(\Theta, \Phi)e^{i\sigma t}.
\label{eq:xir}
\end{equation}
$Y_{\l m}(\Theta,\Phi)$ is the spherical harmonic of degree $\l$ and
order $\m$ in the coordinate system $(\Theta,\Phi)$ aligned with the
pulsation axis, and $\sigma$ is the pulsation frequency.  Compared
with Equation~\ref{eq:dfl}, we have reverted to an expression in which
the imaginary parts of the temporal and spatial dependences are
included; moreover, we have followed Robinson et al.\
(\cite{robikn:82}) in adopting an extra term ${\rm e}^{-i\psi}$ to
allow for the nonadiabatic effects which may introduce a phase
difference between the radial displacement and the flux
changes\footnote{In Robinson et al.\ (\cite{robikn:82}), Equations~20,
23 and~24 should not have negative signs on their right hand sides.
The quantity $\frac{\delta T}{\delta r}$ is positive for adiabatic
pulsations, since maximum radial displacement corresponds to maximum
temperature.  This has no effect on any of the conclusions Robinson et
al.\ presented.}.  The negative sign in the exponent indicates that
for positive values of $\psi$, maximum flux lags maximum radial
displacement.

In Equations~\ref{eq:fl0obs} and~\ref{eq:dflobs}, the value of
$\lambda$ should be the wavelength from which light is Doppler shifted
into $\lobs$, or,
\begin{equation}
\lambda = \lobs\left[1-\frac{v_{\rm rad}(\mu,\phi,t)}{c}\right]
\label{eq:l}
\end{equation}
to first order. We use $v_{\rm rad}$ to express the velocity component
projected into our line of sight.  Calculating $v_{\rm rad}$ requires
first an expression for the pulsation velocities at the stellar
surface.

In the frame of reference $(\Theta,\Phi)$ aligned with the pulsation
axis, the components of the pulsation velocities are, following
Dziembowski (\cite{dzie:77}),
\begin{eqnarray}
V_r&=&i\sigma\epsilon R_0Y_{\ell m}(\Theta, \Phi)e^{i\sigma t}, \nonumber \\
V_\Theta&=&\frac{i\epsilon\vert g\vert}{\sigma} \frac{\partial Y_{\ell m}(\Theta, \Phi)}{\partial \Theta} e^{i\sigma t},\; \rm{and} \nonumber \\
V_\Phi&=&-\frac{i\epsilon\vert g\vert}{\sigma} \frac{m}{\sin \Theta} Y_{\ell m}(\Theta, \Phi) e^{i\sigma t}.
\label{eq:v}
\end{eqnarray}
For {\it g}-mode pulsations, $V_r$ is small compared to the other two
and can be ignored.

To get $v_{\rm rad}$ requires transforming the remaining velocity
expressions into our reference frame and projecting them along our
line of sight.  Then Equations~\ref{eq:flobs} through~\ref{eq:l} are
all we need in principle to calculate colour-dependent pulsation
amplitudes in the presence of non-zero velocities.  In practice, this
would be cumbersome and inefficient for arbitrary inclination and \m,
so we have further simplified the problem by aligning the pulsation
axis with our line-of-sight and holding $m=0$.  Contrary to initial
expectations, this simplification comes at almost no expense; the
results for this case scale easily to arbitrary choices for
inclination and \m.  In the Appendix we demonstrate that this is true
and show how the scaling is done.

This simplification allows us to dispense with the expression for
$V_\Phi$ and to replace $\Theta$ and $\Phi$ in all equations with
$\theta$ and $\phi$ for our reference frame.  The $\phi$ dependences
of $v_{\rm rad}$, $\xi_r$, and $I_\lambda$ also disappear, and the
spherical harmonics reduce to Legendre polynomials.  Finally, to make
our fits independent of the particular choices for stellar radius and
pulsation frequency, we have introduced the fitting parameters:
\begin{eqnarray}
a_T&=&R_0 \left\vert \frac{\delta T}{\delta r}\right\vert\epsilon \;\;\rm{and} \nonumber \\
a_v&=&\frac{\epsilon g}{\sigma}.
\label{eq:amps}
\end{eqnarray}
Thus the final 
expressions
we evaluate numerically are:
\begin{eqnarray}
F_{T_0}(\lobs)&=&2\pi R_0^2 \int_0^1 
I(g, T_0, \mu, \lobs[1-\frac{v_{\rm rad}}{c}])\;\mu\;\d\mu,  \nonumber \\
\Delta F_T(\lobs,t)&=&2\pi R_0^2\; a_T \cos(\sigma t - \psi) \times
\nonumber \\
&&\int_0^1 \left.
\frac{\partial I(g, T, \mu, \lobs[1-\frac{v_{\rm rad}}{c}])}{\partial T}\right|_{T_0} 
P_\l(\mu)\;\mu\;
\d\mu,\nonumber \\
v_{\rm rad}(\mu, t)&=& -a_v \frac{\d P_\l(\mu)}{\d \theta}
 \sin\theta \sin(\sigma t).
\label{eq:fl+dfl}
\end{eqnarray}

Where we have kept only the real parts of the temporal variations. The
free parameters in these equations are the temperature and gravity of
the equilibrium model, $T_0$ and $g$, the (non-adiabatic) phase shift
$\psi$, and the amplitudes given by Equation~\ref{eq:amps}.  The
equilibrium radius, $R_0$, cancels when we calculate the fractional
amplitude.

We have modified the code originally provided by E. L. Robinson to
perform these integrals at a series of time steps covering one
pulsation cycle.  Then we have convolved the output spectra with a
Gaussian to emulate seeing.  Finally, we have calculated the pulsation
amplitude, phase, and mean spectrum at each wavelength.

In Paper~I, we defined the quantity $\Delta\Phi_V$ as the phase
difference between maximum light and maximum velocity.  In the
formalism of this paper,
\begin{equation}
\Delta\Phi_V = \frac{\pi}{2}-\psi.
\label{eq:dv}
\end{equation}
For the $\psi = 0$ case, where flux and radial displacement are in
phase, the $\frac{\pi}{2}$ delay enters because of the time derivative
used to get $v_{\rm rad}$.  Positive values of $\psi$ then reduce
$\Delta\Phi_V$ by delaying the maximum light so that it arrives less
than $\frac{\pi}{2}$ before maximum velocity.  In all of the modes for
which we can detect velocities, $\Delta\Phi_V$ lies in the first
quadrant, implying that flux maximum is delayed compared to the
adiabatic case (or velocity maximum advanced, which is harder to
imagine). This is a profound result, and the first direct
observational constraint on the behaviour of eigenmodes near the
surface of a white dwarf star.

Comparing our models to the data presents something of a challenge,
because of the generally poor match for any choice of amplitude and
phase.  We have simplified the problem by choosing the equilibrium
temperature and gravity from our fit to the average spectrum.  Then we
have fitted $a_T$ directly, by insisting that the fractional amplitude
of the flux changes in our model match those in the data at 5500~\AA.
To choose appropriate values for $a_v$ and $\psi$, we used our model
to calculate several time series of synthetic spectra and then we
reduced them in the same way as the real data.  This allowed us to
calibrate the scaling introduced by the integrals in
Equation~\ref{eq:fl+dfl} for various choices of \l, $a_v$, and $\psi$.
Then we used these scale factors to generate models based on the
properties measured for modes in Paper~I.  The size of the velocities
we use to match our data is quite similar to the size predicted by
Robinson et al.\ (\cite{robikn:82}), $\sim7$~\kms.

\begin{figure}
\plotone{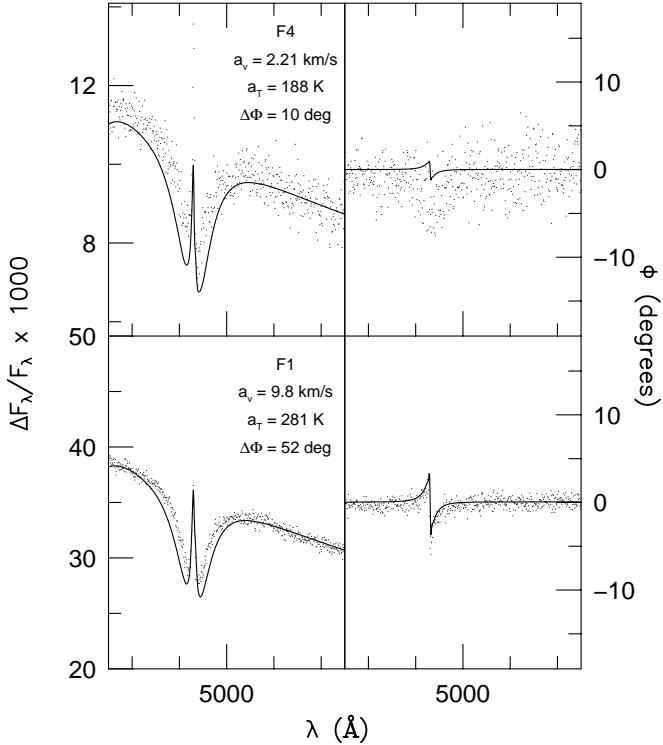}
\caption[]{ Fits to F1 and F4 including velocities. The amplitudes
listed are those described by Equation~\ref{eq:amps} in the text.
\label{fig:profit}}
\end{figure}

Figure~\ref{fig:profit} shows the results of these calculations for
the H$\beta$ line in modes F1 and F4.  Our models now reproduce the
asymmetry in the amplitude plots, and show phase changes within the
line profiles.  These phase changes could never be reproduced without
including velocities in the models.  However, the overall quality of
the amplitude fits within lines is not much improved, and is not
changed at all in the continuum.  The phases are a good fit within
spectral lines for the larger modes, but our models still do not
reproduce the slow change in phase observed in the continuum (see
Figure~\ref{fig:phase}).  For now, this remains a mystery, but
probably a mystery hiding interesting pulsation physics.

Finally, we consider the behaviour of the ratio between the observed
velocity and flux amplitudes.  As we show in the appendix, this ratio
does not depend on the inclination or \m\ of the modes, and therefore
cannot be used to constrain those quantities.  However, it is
sensitive to the value of~\l\, as evident from the different forms for
flux and velocity in the integrals of Equation~\ref{eq:fl+dfl}.  The
flux depends on the Legendre polynomial while the velocity depends on
its derivative and in addition is weighted towards the limb by the
projection onto the line of sight.  According to our models, for a
fixed value of $\frac{a_v}{a_T}$, an $\l = 2$ mode should have
$\sim\!4$ times higher apparent velocity to light ratio than an $\l =
1$ mode.  It is then no surprise that mode F4, which we have
identified as $\l = 2$, has the highest velocity to light amplitude
ratio (see Paper~I) of any of the modes we have detected, a result
which supports our identification of that mode as $\l = 2$.

\section{Summary and Conclusions}\label{sec:conc}

  Asteroseismology of ZZ~Ceti stars has been impeded for almost two
decades by the lack of a reliable method for mode identification.  We
have tested a new method that uses high signal to noise time-resolved
spectroscopy to measure the wavelength dependence of optical pulsation
amplitudes.  We have found that the \l-dependent changes in amplitude
predicted by the models are also present in our data, allowing us to
assign values of \l\ to six modes in the star \zzpsc.

 Initially, this will be of greatest benefit to seismological models
of \zzpsc.  None of the modes we have identified is short enough to
yield immediate constraints on the mass of the H layer, as is possible
for \G117-B15A (Robinson et al.~\cite{robi&a:95}), \GD165 (Bergeron et
al.~\cite{berg&a:93}), and \G226-29 (Fontaine et
al.~\cite{font&a:92}).  However, a concerted effort to match the six
modes of known \l\ may yield a unique solution.  If not, it is
possible to return to this star with the hope that four more hours of
data will allow identification of a different set of modes.  Sooner or
later, we will have a definitive asteroseismological solution for this
star; the main obstacle has been removed.

Our method can also be extended to other ZZ~Ceti stars, although the
fainter, low amplitude stars will require longer runs.  We have
already begun a programme to identify \l\ in as many stars as is
practical in the observing time available to us.  This programme
should allow us to measure structural properties of enough stars to
answer some long-standing questions about the DA stars, such as the
masses of the H and He surface layers, which have been the subject of
some controversy (Shipman~\cite{ship:96}; Fontaine \&
Wesemael~\cite{font&w:96}).

Apart from the impact our results will have on ZZ~Ceti seismology,
they have also opened new windows into the surface physics of
pulsating white dwarfs.  We have measured two new diagnostics of the
behaviour of pulsations near the photosphere: the amplitude of the
pulsation velocities, and the phase lag between flux maximum and
velocity maximum.  These can be compared quantitatively with
predictions of nonadiabatic pulsation theories (e.g., Lee \& Bradley
\cite{leeb:93}; Wu \& Goldreich \cite{wug:99}), as well as with
measurements for other stars.  We have already detected velocity
changes in one other star (HL~Tau~76).

Finally, our high signal to noise measurements provide a serious
challenge to the atmospheric model fits.  There are significant
discrepancies between the models and our average spectrum and between
the models and the spectral changes caused by pulsation.  We have not
identified the source of these differences, but hopefully our data and
future data like them will provide assistance to modelers in their
continuing attempts to understand the atmospheres of white dwarf
stars.

\acknowledgements This paper relied upon many enlightening discussions
with Scot Kleinman, Kepler Oliveira, Rob Robinson, Peter Goldreich,
and Mike Montgomery.  We are grateful to each of them.  We also thank
the referee, Detlev Koester, for useful comments.  J.C.C.  and his
family are grateful to the Sherman Fairchild Foundation for providing
the support that made it possible to carry out this research.
M.H.v.K.\ acknowledges a NASA Hubble Fellowship while at Caltech, and
a fellowship of the Royal Netherlands Academy of Arts and Sciences at
Utrecht.  The observations reported here were obtained at the
W.~M.~Keck Observatory, which is operated by the California
Association for Research in Astronomy, a scientific partnership among
the California Institute of Technology, the University of California,
and the National Aeronautics and Space Administration.  It was made
possible by the generous financial support of the W.~M.~Keck
foundation.

\appendix
\section{Arbitrary inclination and value of \m}\label{sec:general}
  The usual way to transform an arbitrary spherical harmonic into the
coordinate system aligned with our line of sight is to recognise that
the $2\l+1$ spherical harmonics of a given \l\ form a complete basis
set, so it is always possible to express an arbitrary spherical
harmonic as a sum of spherical harmonics in our coordinate system,
\begin{equation}
Y_{\l m}(\Theta, \Phi) = \sum_{m^\prime = - \l}^\l R_{m^\prime
m}^\l Y_{\l m^\prime}(\theta, \phi),
\label{eq:ylm}
\end{equation}
The coefficients $R_{m^\prime m}^\l$ are a function of the inclination
angle (see Dziembowski \cite{dzie:77}; Robinson et al.\
\cite{robikn:82}).  Then the $m^\prime \neq 0$ terms of the sum cancel
to zero in any integration over the visible hemisphere, so that only
the $m^\prime = 0$ term aligned with our line of sight remains.

  This will not work directly for the expressions we integrate in
Equations~\ref{eq:fl0obs} and~\ref{eq:dflobs}, because they contain
$I(g, T, \mu, \lambda)$ and its derivative with respect to $T$, which
may be arbitrarily complex functions of the eigenmode velocities.
However, if we expand $I$ in a Taylor series about $\lambda$,
\begin{eqnarray}
I(g, T, \mu, \lambda)&=&I(g, T, \mu, \lobs) \nonumber \\
&+&\left.\frac{\partial I(g,T,\mu, \lambda)}{\partial
\lambda}\right|_{\lobs}(\lambda-\lobs),
\label{eq:il}
\end{eqnarray}
do the same for $\frac{\partial I}{\partial T}$, and use
Equation~\ref{eq:l}, our expression for the time dependent flux
becomes
\begin{eqnarray}
F(\lobs,t)&=& 2\pi R_0^2\int_0^1 I(\mu,\lobs)\;\mu\; \d\mu\ \nonumber \\
&-& R_0^2\frac{\lobs}{c}\int_0^1\int_0^{2\pi} v_{\rm rad}\;
\d\phi\;\left.\frac{\partial I(\mu,\lambda)}{\partial
\lambda}\right|_{\lobs}\;\mu\; \d\mu\ \nonumber \\
&+& R_0^2 a_T {\rm e}^{i\sigma t} {\rm e}^{-i\psi}\times\nonumber \\
&&\int_0^1\int_0^{2\pi} Y_{\l m}\; \d\phi\;
\left.\frac{\partial I(\mu,\lobs)}{\partial T}\right|_{T_0}
\mu\; \d\mu.       
\label{eq:flobsapp}
\end{eqnarray}
We have suppressed the g and T dependencies, which vanish upon
choosing an equilibrium model.  We have also left out the cross term
which includes the product of $\frac{\partial I}{\partial T}$ and
$\frac{\partial I}{\partial \lambda}$ because these are both small
quantities.  The remaining expression contains separate terms for the
equilibrium flux ($F_0$), the flux changes due to velocity shifts
($\Delta F_v$), and the flux changes due to temperature
changes($\Delta F_T$).  Using Equation~\ref{eq:ylm}, we can express
both $v_{\rm rad}$ and $Y_{\l m}$ in the above integral into the
coordinate system $(\theta,\phi)$; for both, the terms with
$m^\prime\neq0$ integrate to zero in the $\phi$ direction (Dziembowski
\cite{dzie:77}), so only the $m = 0$ term aligned to our line of sight
remains.  Thus we may write:
\begin{eqnarray}
F(\lobs,t)&=&F_0(\lobs) \nonumber \\
&+& R_{0 m}^\l \Delta F_v(\lobs,t,i=0,m=0) \nonumber \\
&+& R_{0 m}^\l \Delta F_T(\lobs,t,i=0,m=0).
\label{eq:flobsapp2}
\end{eqnarray}

This expression permits us to relate $\Delta F/F (\lobs, i\neq 0,
m\neq 0)$ to $\Delta F/F (\lobs, i = 0, m = 0)$ via a simple scaling
factor $R_{0 m}^\l$.  As $R_{0 m}^\l$ does not depend on $\lobs$, the
results shown in Figure 3 are general for arbitrary inclination and
$m$ value.  Also, because the factor is the same for both the
temperature and velocity induced flux changes, the observed ratio
$a_v/a_T$ does not depend on $i$ and $m$.  Finally,
Equation~\ref{eq:flobsapp2} shows that the line-of-sight velocities
and flux variations are separable to first order, a result used
implicitly in Paper~I.

It is important to recognise where these approximations may break
down.  When the second derivative of the spectrum dominates the first,
which can happen in the central few Angstroms of an absorption line,
our expansion to first order in $\lambda$ is not reliable.

\begin{thebibliography}{}
\bibitem[1992]{bergsl:92}
   Bergeron, P., Saffer, R. A., \& Liebert, J. 1992, ApJ, 394, 228
\bibitem[1993]{berg&a:93}
   Bergeron, P., et al. 1993, AJ, 106, 1987
\bibitem[1995]{berg&a:95}
   Bergeron, P., Wesemael, F., Lamontagne, R., Fontaine, G., Saffer,
   R. A., \& Allard, N. F.  1995, ApJ, 449, 258
\bibitem[1996]{bradk:96}
   Bradley, P. \& Kleinman, S. 1996, in "White Dwarfs", ed.\ J.~Isern,
M.~Hernanz, \&
\bibitem[1995]{brasfw:95}
   Brassard, P., Fontaine, G., \& Wesemael, F. 1995, ApJS, 96 545
  E.~Garcia-Berro (Dordrecht: Kluwer), 445
\bibitem[1992b]{bric:92}
   Brickhill, A. J.  1992, MNRAS, 259, 529
\bibitem[1977]{dzie:77}
    Dziembowski, W. 1997, Acta Astron, 27, 203
\bibitem[1997]{finlkb:97}
   Finley, D. S., Koester, D. \& Basri, G.  1997, ApJ, 488, 375 
\bibitem[1992]{font&a:92}
    Fontaine, G., Brassard, P., Bergeron, P. \& Wesemael, F. 1992, ApJ,
399, L91
\bibitem[1996]{font&w:96}
   Fontaine, G., \& Wesemael, F. 1996, in "White Dwarfs", ed.\ J.~Isern,
M.~Hernanz, \&
   E.~Garcia-Berro (Dordrecht: Kluwer), 173
\bibitem[1997]{font&a:97}
    Fontaine, G., Brassard, P., Bergeron, P. \& Wesemael, F. 1997, ApJ,
469, 320
\bibitem[1984]{kepl:84}
   Kepler, S. O. 1984, ApJ, 286, 314
\bibitem[1995]{klei:95}
   Kleinman, S. J.  1995, PhD thesis, University of Texas at Austin
\bibitem[1994]{klei&a:94}
   Kleinman, S. J., et al.  1994, ApJ, 436, 875
\bibitem[1998]{klei&a:98}
   Kleinman, S. J., et al.  1998, ApJ, 495, 424
\bibitem[1994]{koesav:94}
   Koester, D., Allard, N. F., \& Vauclair, G. 1994, A\&A, 291, L9
\bibitem[1997]{koesps:97}
   Koester, D., Provencal, J., \& Shipman, H. L.  1997, A\&A, 230,
   L57
\bibitem[1999]{leeb:93} 
  Lee, U., \& Bradley, P. A.  1993, ApJ, 418, 855
\bibitem[1995]{oke&a:95}
   Oke, J. B., et al.  1995, PASP, 107, 375
\bibitem[1982]{robikn:82}
   Robinson, E. L., Kepler, S. O., \& Nather, R. E.  1982, ApJ, 259, 219

\bibitem[1995]{robi&a:95}
   Robinson, E. L., et al.  1995, ApJ, 438, 908
\bibitem[1996]{ship:96}
   Shipman, H. L. 1996, in "White Dwarfs", ed.\ J.~Isern, M.~Hernanz, \&
   E.~Garcia-Berro (Dordrecht: Kluwer), 165
\bibitem[1999]{vank&a:99}
   Van Kerkwijk, M. H., Clemens, J. C., \& Wu, Y.  1999, MNRAS,
   accepted (Paper~I) 
\bibitem[1990]{wing&a:90}
   Winget, D. E., et al.  1990, ApJ, 357, 630
\bibitem[1991]{wing&a:91}
   Winget, D. E., et al.  1991, ApJ, 378, 326
\bibitem[1994]{wing&a:94}
   Winget, D. E., et al.  1994, ApJ, 430, 839
\bibitem[1994]{wood:94}
   Wood, M. A., 1994, BAAS, 185, 4601
\bibitem[1999]{wug:99}
   Wu, Y., \& Goldreich, P. 1999, ApJ, 519, 783
\bibitem[1987]{zuckb:87}
   Zuckerman, B., \& Becklin, E. E.  1987, Nature, 330, 138
\end{thebibliography}
\end{document}